\documentclass[11pt]{article}
\usepackage{graphicx}
\usepackage{amssymb}

\begin{document}
\title{Event Weighted Tests for Detecting Periodicity in Photon Arrival Times}
\author{Peter Bickel\footnotemark[1], Bas Kleijn\footnotemark[2], \& John Rice\footnotemark[1]}
\renewcommand{\thefootnote}{\fnsymbol{footnote}}
\footnotetext[1]{Department of Statistics, University of
California, Berkeley} %
\footnotetext[2]{Korteweg-de Vries Institute for Mathematics,
       Faculty of Science, University of Amsterdam}
\maketitle

\begin{abstract}
This paper treats the problem of detecting periodicity in a sequence of photon arrival times, which occurs, for example,  in attempting to detect gamma-ray pulsars. A particular focus is on how auxiliary information, typically source intensity, background intensity,  and incidence angles and energies associated with each photon arrival should be used to maximize the detection power. We construct a class of likelihood-based tests, score tests, which give rise to event weighting in a principled and natural way, and derive expressions quantifying the power of the tests.  These results can be used to compare the efficacies of different weight functions, including cuts in energy and incidence angle.  The test is targeted toward a template for the periodic lightcurve, and we quantify how deviation from that template affects the power of detection.
\end{abstract}

\section{Introduction}
From a sequence of photon arrival times $0 \leq t_1 < t_2 < \cdots t_N < T$, we wish to test the hypothesis that some of the photons come from a periodic source (for example, a gamma-ray pulsar) versus the null hypothesis that they  come from a background plus a source that does not vary in time. The background emission rate is assumed to be constant in time. Associated with each event is  auxiliary information, such as the incidence angle and the measured energy; we denote these variables by $z$.   
Ignoring this information is clearly wasteful, and in fact it would typically be used, at least in the form of cuts in energy and incidence angle.  The value of $z$ associated with an event (an arrival) provides information about the relative likelihood that photon was from the source or the background, and it seems intuitively  that the event should be correspondingly weighted in some manner. (Note that  cuts corresponds to weights that are zero or one.)
 A main thrust of this paper is to derive in a principled way how this information can best be used to enhance detection power.  We derive expressions which quantify the efficiency of any weighting function and the form of the optimal function.

Unless the periodic light curve is known, there is no universally optimal test, since a test that is most powerful against one light curve will not be most powerful against another.  This statement also applies to tests that attempt to adapt to the form of the lightcurve. Any test implicitly or explicitly  commits to a  finite dimensional class of targets. Generally, the light curve of the source is unknown, so we consider testing against a template, a probability density function $\nu_0(t)$ on $[0,1]$, extended periodically, with Fourier series
\begin{equation}
\nu_0(t) = 1 + \eta \sum_{n \neq 0} \alpha_n e^{2 \pi i n t}
\end{equation}
for  $\eta \geq 0$. If $\eta=0$, the source intensity is constant in time.
Defining $\nu_\tau(t) = \nu_0(t + \tau)$
\begin{equation}
\nu_\tau(t) =  1 + \eta \sum_{n \neq 0} \alpha_n e^{2 \pi i n t + 2 \pi i n \tau},
\end{equation}

We  model the arrival times as  the superposition of independent background and source processes,  a Poisson process with rate function
\begin{equation}
\lambda(t | \theta, \tau, \mu, f) = \mu c(t) [(1-\theta) + \theta
\nu_\tau(\phi(t))], ~~ 0 \leq \theta \leq 1
\end{equation}
where $c(t)$ denotes the sensitivity of the instrument at time $t$.
Here  $\theta$ is the proportion of flux from source; the phase function is $\phi(t) = ft$, or if drift is taken into consideration, $\phi(t) = ft + \dot{f}t^2/2$.
Within this framework, different hypotheses can be tested. We focus on testing the null hypothesis $H: \eta=0$ versus the alternative $K: \eta > 0$.  That is, we are concerned with a situation in which the presence of a source is not in doubt, but its periodicity is in question.  Testing whether there is any source at all corresponds to testing $H_2: \theta=0$ against $K_2: \theta> 0$.  

This  paper extends some results of Bickel et al.~(2007), with more extensive considerations of  event weighting. In the next section we derive a test which makes use of the information contained in both the arrival times, $t_j$,  and the associated variables, $z_j$, in a principled way, by appropriately weighting the arrival times.  In Section 3, we show how the detection power of the test depends on the weights.  Expressions derived there allow comparison of power when ideal weights are used and using approximate weights, such as simple cuts. We will also see the price paid for mismatch between the template and the actual light curve and for mismatch of the specified frequency and the actual frequency.  Section 4 contains some illustrative examples.  Some technical details are deferred to an Appendix.

\section{Score test}
Let $f_B(z)$ denote the probability density function of $z$ for a background event and $f_S(z)$ the density function for a source event. We base a test on the likelihood function, assuming  that the $z_j$ are independent of the arrival times:
\begin{equation}
\label{eqn-Lik}
L = \mu^N \prod_{j=1}^N c(t_j)[(1- \theta) f_B(z_j) + \theta f_S(z_j) \nu_\tau(\phi(t_j))]
\exp \Big(- \mu \int_0^T c(t)[ (1-\theta) + \theta \nu_\tau(\phi(t))] dt \Big)
\end{equation}
A score test (Lehman and Romano, 2006)  of \emph{H} versus \emph{K} is formed by evaluating the derivative of the log likelihood at $\eta=0$:
\begin{equation}
\label{score}
S(\tau) = \sum_{j=1}^n \left(
\frac{ \theta f_S(z_j)}{(1-\theta) f_B(z_j) + \theta f_S(z_j)} (\nu_\tau(\phi(t_j))-1) \right)  - \mu \theta
\int_0^T c(t)[\nu_\tau(\phi(t))-1]dt \label{score-tau}
\end{equation}
If  $\phi(T) \gg 1$  and $c(t)$ varies slowly and is nonzero over a substantial fraction of $[0,T]$, the second term is neglible.  We will make this assumption throughout.  

Let
\begin{equation}
\label{wj}
w_j = \frac{ \theta f_S(z_j)}{(1-\theta) f_B(z_j) + \theta f_S(z_j)}
\end{equation}
This is the probability that photon $j$ is from the source, given $z_j$. For a very weak source (small $\theta$), an approximation to (\ref{wj}) gives $w_j \propto f_S(z_j)/f_B(z_j)$.
If $z = (E, \varphi)$, energy and incidence angle, we can write
\begin{eqnarray}
f_B(z) & = & f_B(E) f_B(\varphi |E) \\
f_S(z) & = & f_S(E) f_S(\varphi |E) \\
w(z) & = & w(E) w(\varphi |E) 
\end{eqnarray}
The optimal weight function is then
\begin{equation}
\label{}
w_{\mbox{opt}}(E, \varphi) = \frac{ \theta f_S(E) f_S(\varphi |E)}{
\theta f_S(E) f_S(\varphi |E) + (1-\theta) f_B(E) f_B(\varphi |E)}
\end{equation}
For a weak source (small $\theta$), we have the approximation
\begin{equation}
\label{}
w_{\mbox{opt}} \propto \frac{f_S(E)f_S(\varphi |E)}{f_B(E)f_B(\varphi | E)} 
\end{equation}
The function $f_S(\varphi |E)$ is the point spread function of incidence angle at energy $E$.  The background would normally be assumed to be spatially uniform, from which $f_B(\varphi |E)$ would follow.  The optimal weight function also
depends on the ratio of the energy spectra of source to background,  which potentially provides valuable information, but might be unknown in practice. In the latter case one could use a weight function,
\begin{equation}
\label{ }
w(E,\varphi) =  \frac{ \theta  f_S(\varphi |E)}{
\theta  f_S(\varphi |E) + (1-\theta)  f_B(\varphi |E)}
\end{equation}
or for small $\theta$
\begin{equation}
\label{}
w(E, \varphi) = \frac{ f_S(\varphi |E)}{ f_B(\varphi |E)}
\end{equation}

The test statistic (\ref{score}) depends on the data through
\begin{eqnarray}
\sum_{j=1}^N w_j (\nu_\tau(\phi(t_j)-1) & = & \sum_{j=1}^N w_j \sum_{n \neq 0} \alpha_n e^{2 \pi i n \phi(t_j) + 2 \pi i n \tau} \\
& = & \sum_{n \neq 0} \alpha_n A_n e^{2 \pi i n \tau}
\end{eqnarray}
where $A_n = \sum_j w_j \exp(2 \pi i n \phi(t_j))$.   To eliminate the dependence of the test statistic on the phase, $\tau$, we use $ \int_0^1 |S(\tau)|^2 d\tau$.  By Parseval's theorem
\begin{equation}
\label{}
\int_0^1 \Big| \sum_{n \neq 0} \alpha_n A_n e^{2 \pi i n \tau} \Big|^2 d\tau = \sum_{n \neq 0} |\alpha_n|^2 |A_n|^2
\end{equation}
Unless the source is weak, the statistic depends upon $\theta$, which may be approximately known from other analyses, or may be estimated by maximum likelihood under the null hypothesis. In the latter case, the log likelihood is
\begin{eqnarray}
\label{}
\ell(\theta) & = & N \log \mu + \sum_{j=1}^N \log c(t_j)  + \nonumber \\
& & \sum_{j=1}^N \log[(1-\theta) f_B(z_j) + \theta f_S(z_j)] - \mu \int_0^T c(t)dt.
\end{eqnarray}
The log likelihood depends on $\theta$ only through the third term, which can be easily maximized numerically, if $f_B(z)$ and $f_S(z)$ are known.  The final test statistic either uses the value of $\theta$ known \emph{a priori} or the maximum likelihood estimate:
\begin{equation}
Q_T = \frac{1}{T} \sum_{n \neq 0}|\alpha_n|^2 |A_n|^2
\end{equation}

The score test is an attractive alternative to a  generalized likelihood ratio test.  To compute the likelihood ratio test, the likelihood (\ref{eqn-Lik}) would have to be maximized both under $H$ and $K$, and the latter would entail estimating the parameters $\theta$, $\eta$ and $\tau$.

Beran (1969) showed that this test, in an unweighted form, was locally most powerful invariant for testing uniformity of a distribution on the circle.
In the case $|\alpha_n| = 0, ~ n > 1$ and $w_j =1$, this is Rayleigh's test (Rayleigh, 1919).  If $|\alpha_n| = 1, ~ n \leq m$ and $|\alpha_n| = 0, ~ n > m$ and $w_j=1$, this is the $Z_m^2$ test of Buccheri et al.~(1983). De Jager et al.~(1989) proposed the $H$-test, which chooses $m$ adaptively.   

We now consider the  distribution  of $Q_T$ when there is no periodicity  ($\eta = 0$).  Let $\beta_1 = \int w(z) f_B(z)dz$ and $\zeta_1 = \int w(z) f_S(z)dz$  be the expected values of the weight of  background and source events and let $\beta_2 = \int w^2(z) f_B(z)dz$ and $\zeta_2 = \int w^2(z) f_S(z) dz$.   The average value of a weight is $E(W) = (1-\theta)\beta_1 + \theta \zeta_1$ and $E(W^2) = (1 - \theta)\beta_2 + \theta \zeta_2$.  Let $\mu_0 = \mu T^{-1} \int_0^T c(t)dt$. In the Appendix we argue that
\begin{eqnarray}
E_H Q_T  & \simeq & [(1-\theta) \beta_2 + \theta \zeta_2] \mu_0  \sum_{n \neq 0} |\alpha_n|^2 \\
Var_H(Q_T) & \simeq  &  [(1-\theta) \beta_2 + \theta \zeta_2]^2\mu_0^2 \sum_{n \neq 0} |\alpha_n|^4
\end{eqnarray}
Also $2 |A_n|^2/(\mu_0 T [(1-\theta) \beta_2 + \theta \zeta_2] )$ has approximately a chi-square distribution with two degrees of freedom.
The $A_n$ are approximately independent so that $Q_T$  approximately has the distribution of a weighted sum of independent chi-square random variables.  The scaling of the chi-square random variables can be estimated as follows: observe that since  $\mu_0T$ is the expected number of events in $[0,T]$, $\sum w_j^2 \simeq \mu_0T E(W^2)$.  Thus
\begin{equation}
\label{ }
\mu_0T [(1-\theta) \beta_2 + \theta \zeta_2] \simeq \sum_j w_j^2
\end{equation}

\section{Power}

We next consider properties of the test statistic  $Q_T$ when there is a periodic source, i.e. $ \eta > 0$.  Let the pulse shape of the source be 
\begin{equation}
\label{gamma_n}
\gamma(t) = \sum_{n \neq 0} \gamma_ne^{2 \pi i n t}
\end{equation}

As an indication of the detection power of the test, we can use the signal to noise ratio.  Let $E_H(Q_T)$ and $E_K(Q_T)$ respectively denote the expected values of the test statistic $Q_T$ when there is and is not a periodic component, and let $\sigma_H$ denote the standard deviation of $Q_T$ under the null hypothesis of no periodic component.    If the phase function $\phi(t)$ is correctly identified (e.g. if $f$ and $\dot{f}$ are correctly specified) we show in the Appendix that
\begin{equation}
\label{SNR}
\frac{E_K(Q_T) - E_H(Q_T)}{\sigma_H} \simeq
\theta^2 T \mu_0 \mathcal{E}(w) \frac{  \sum_{n \neq 0} 
|\gamma_n|^2 |\alpha_n|^2}{ [\sum_{n \neq 0} |\alpha_n|^4 ]^{1/2}}
\end{equation}

Here the efficiency of the weighting function enters as
\begin{equation}
\label{EffW}
\mathcal{E}(w)  =  \frac{\zeta_1^2}{(1-\theta) \beta_2 + \theta \zeta_2} 
 =  \frac{[E(W | ~ \mbox{source})]^2}{E(W^2)} \label{Eff}
\end{equation}
This expression holds for any weight function. Since a weight function need only be defined up to a constant of proportionality, the denominator provides a normalization.  The optimal weight function is that given by the score test, (\ref{wj}), in which case it follows from a short calculation that
\begin{equation}
\label{Eopt}
\mathcal{E}(w_{\mbox{opt}}) = \frac{1}{\theta} \int \frac{ \theta f_S(z)}{(1-\theta)f_B(z) + \theta f_S(z)} f_S(z) dz
\end{equation}
which is the ratio of the average  probability of a source event given $z$ to the marginal probability of a source event. The efficacy of weighting depends in this way on the degree to which $z$ discriminates between background and source, or on how correlated it is with the optimal weight function,  
 since after some algebra,
\begin{equation}
\label{}
\mathcal{E}(w) = \frac{[E(WW_{\mbox{opt}})]^2}{E(W^2)}
\end{equation}
where the expectations are taken with respect to the marginal density of $Z$, $(1-\theta)f_B(z) + \theta f_S(z)$. 
In the case of no weighting, $w(z)=1$,
$\mathcal{E}=1$. 

From (\ref{SNR}), the detection threshold for a weak signal is $\theta$ of the order  $T^{-1/2}$.  The expression also quantifies how the power depends upon the match of the template $\{ |\alpha_n|^2 \}$ to the source profile $\{ |\gamma_n|^2 \}$. Maximal power is achieved when $|\alpha_n|^2 \propto |\gamma_n|^2$. So for detection of periodicity of a given source, the best
detection-statistic has the same spectrum as the source in question.
Because the latter is unknown,  a template could be based on known sources (see Section 4 for an example).

This result assumes that $\phi(t)$ is very accurately specified.  In the case $\phi(t) = f_0 t + \dot{f_0}t^2/2$, and approximate values are used,
$f = f_0 + \Delta/T$ and $\dot{f} = \dot{f_0} + \Delta/T^2$, $\Delta < 1$, the sum in the numerator of (\ref{SNR}) becomes
$\sum_{n \neq 0} |\gamma_n|^2 |\alpha_n|^2 ( 1 - O((n\Delta)^2))$. Thus, accurate specification is especially important for higher harmonics to contribute to the power.  This depends on the rate of decay of $\gamma_n$ and on that of $\alpha_n$, which for practical reasons would be zero for sufficiently large $n$.

\section{Examples}
\subsection{Template}
To illustrate the effect of the template choice,  $\{|\alpha_n|^2\}$, we phased photon arrival times from Crab, Geminga, and Vela for single EGRET viewing periods. We calculated the corresponding coefficients, $|A_n|^2$ (with no weighting).  For pedagogical illustration, we normalized them and regard them as the coefficients  $|\gamma_n|^2$ (\ref{gamma_n}) of the sources. 
These coefficients are plotted in Figure 1.  It is interesting that in all cases the coefficient $|\gamma_2|^2$ is largest. 

\begin{figure}
\begin{center}
\includegraphics[scale=0.5]{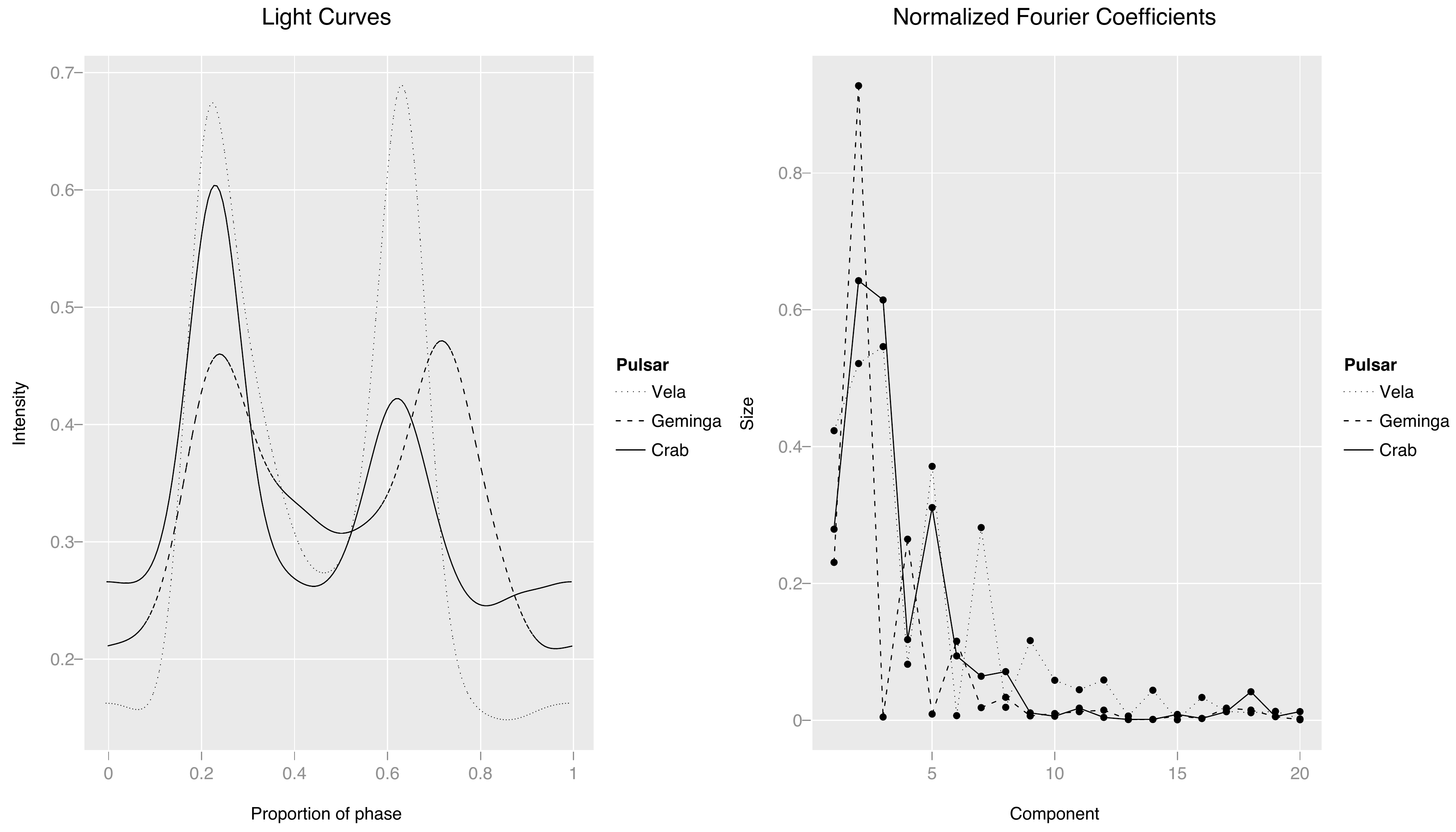}
\caption{Left panel: smoothed light curves, computed from single EGRET viewing periods, for the  Crab, Geminga, and Vela pulsars. Right panel: normalized coefficients $|A_n|^2$ for each pulsar.}
\label{Famps}
\end{center}
\end{figure}

The template  $\{|\alpha_n|^2\}$ will be most powerful for a particular source if $|\alpha_n|^2 \propto |\gamma_n|^2$, which is of course not possible in practice.  To illustrate the effects of suboptimal templates, we evaluated percent efficiency for sequences $|\alpha_n| = 1$, $n \leq m$ and $|\alpha_n | = 0$, $n > m$, for $m=1,2,\ldots, 10$ (the $Z_m^2$ test).   (By ``efficiency'' we mean the percentage of the signal to noise ratio (\ref{SNR}) that is attained relative to that attained by the optimal template, $|\alpha_n|^2 \propto |\gamma_n|^2$.)  The results are displayed in Table 1. As would be expected from Figure 1, the efficiency increases initially with $m$, and then decreases.  Considerable gains in power would result from using two to five harmonics, since the signal to noise ratios increase  by factors of two to  three.  For example, one would expect that a 8.9$\sigma$ result using the first three coefficients for Crab would only be a 2.8$\sigma$ result using the Rayleigh test.  We also experimented with using the average of the three sources as a template, cutting off after five and ten terms.  Those results are shown in Table 2. (The first five average coefficients are 0.35, 0.77, 0.43, 0.17, 0.26.) Very little is gained in going from five to ten non-zero coefficients, and the computational savings would be substantial, since we would need $|n \Delta| < 1$ for the highest harmonic.  For example, if one were using ten harmonics the "natural" Fourier frequencies $k/T$ would have to be oversampled by a factor of at least ten and all ten harmonics would have to be calculated.

\begin{table}[htdp]
\caption{Relative efficiencies for $m=1,2, \ldots, 10$.}
\begin{center}
\begin{tabular}{|c|c|c|c|c|c|c|c|c|c|c|}
\hline
number of coefficients & 1 & 2 & 3& 4&5&6&7&8&9&10 \\
\hline
Crab & 28& 65& 89&83 &88 &84 &80 &78 &74 & 70 \\
\hline
Geminga &23 &82 & 67& 71& 66& 63& 59& 57&54 & 52 \\
\hline
Vela & 42  & 67 & 86 & 79 & 87 & 80 & 84 &79 &79 & 77 \\
\hline
\end{tabular}
\end{center}
\label{default}
\end{table}%

\begin{table}[htdp]
\caption{Relative efficiencies obtained from using the first five and first ten average coefficients as the template.}
\begin{center}
\begin{tabular}{|c|c|c|}
\hline
number of terms & 5 & 10 \\
\hline
Crab & 96 & 97 \\
\hline
Geminga & 85 & 85 \\
\hline
Vela & 89 & 93 \\
\hline
\end{tabular}
\end{center}
\label{default}
\end{table}%

\subsection{Weight function}
 
Consider a source which emits photons at rate $\alpha$ and a background whose rate is $\rho$ per unit area  and suppose that photons are collected in a disc of radius $R$ (rather than a spherical cap, for simplicity).
Then
\begin{eqnarray}
\mu & = & \pi R^2 \rho + \alpha \\
\theta & = & \frac{\alpha}{\pi R^2 \rho + \alpha} \\
f_B(\varphi |E) & = & \frac {2 \varphi}{R^2}, ~~ 0 \leq \varphi \leq R
\end{eqnarray}
Let $\beta =  2 \pi \rho/\alpha$, a measure of the strength of the background relative to the source. Then the denominator of (\ref{Eff}) is
\begin{eqnarray}
\label{ }
E(W^2) &  = & \frac{\alpha}{\pi R^2 \rho + \alpha} \Big[
\int w^2(E) f_S(E) \int w^2(\varphi|E) f_S(\varphi|E) d\varphi dE + \nonumber \\
& & \beta \int w^2(E) f_B(E) \int \varphi w(\varphi|E) d\varphi dE \Big]
\end{eqnarray}
The factor $\alpha/(\pi R^2 \rho + \alpha)$ when combined with the factor
$\theta^2 \mu_0$ in (\ref{SNR}) is proportional to $\alpha$.
The optimal weight function is then
\begin{equation}
\label{}
w_{\mbox{opt}}(E, \varphi) = \frac{f_S(E) f_S(\varphi |E)}{
f_S(E) f_S(\varphi |E) + \beta \varphi f_B(E)}
\end{equation}
which depends on the energy spectra through their ratio. 

If the psf is bivariate circular Gaussian with standard deviation $\sigma(E)$, then
$\varphi$, the distance to the origin, has the probability density function
\begin{equation}\label{}
    f_S(\varphi |E) = \frac{\varphi}{\sigma(E)} \exp(-
    \frac{\varphi^2}{2 \sigma^2(E)})
\end{equation}
(This assumes that $\sigma(E) \ll R$, otherwise the density truncated at $R$ has to be normalized to have unit area.)
Then the optimal weight function is 
\begin{equation}
\label{}
w_{\mbox{opt}}(E, \varphi) = \frac{f_S(E) 
}{
f_S(E)  + \beta  \sigma(E) \exp(\varphi^2/2 \sigma^2(E)) f_B(E)}
\end{equation}
 If photons are not differentially weighted according to the ratio of the energy spectra, one has the weight function
 \begin{equation}
\label{}
w(E,\varphi) = \frac{1
}{
1  + \beta  \sigma(E) \exp(\varphi^2 /2 \sigma^2(E)) }
\end{equation}
The decay of this weight function depends on the parameter $\xi=\beta \sigma(E)$.  If this parameter is very large  (weak source/strong background/large $ \sigma$), $w(\varphi ) \propto \exp( - (\varphi^2/2 \sigma^2(E)))$.  Numerical exploration shows that there is little difference among the functions for $\xi \geq 1$, but if $\xi=0.1$ and $\xi=0.01$, the weights decay substantially more slowly. For example, if $\xi=1$ a 2$\sigma$ incidence angle is given weight (relative to that of a photon that is directly on source) of about 0.1 and a 3$\sigma$ angle is given approximately zero weight.  In comparison, for $\xi=0.01$ a 2$\sigma$ angle receives weight about 0.9, a 3$\sigma$ angle receives weight about 0.5 and a 5$\sigma$ angle receives weight approximately 0.

\section{Conclusion and Discussion}
We have presented a class of tests that depend on two features: a template for the form of the periodic light curve and a function that differentially weights arrival times. We have suggested using a template constructed as the average of those of known sources, but one could choose the template adaptively, for example by considering the maximum of the test statistic over the light curves from the known sources.
The power of such a test would be more difficult to analyze explicitly, as would be the power of the $H$-test.                                                                                                                                                                                          From general theory we know that any test will not be uniformly most powerful, but will perform better in certain ``directions'' than others.  Janssen (2000) shows that in testing for uniformity any test can achieve high power for at most a finite dimensional family of alternatives.  This can be seen quite clearly in  expressions we have developed to quantify the power of the test (\ref{SNR}). 

Ideally, the weight given to a photon arrival should be proportional to the probability that the photon came from the source, given its measured energy, incidence angle, and any other available information. The optimal weight function  can only be approximated in practice.  It depends on the ratio of the energy spectra of the source and background, which may not be accurately known for a faint source.  It depends, through $f_S(\varphi|E)$, on the source location, which may also be subject to uncertainty.  The efficiency of any weight function, $w(z)$,  has the conceptually simple form (\ref{EffW}).  

The score test was derived under some assumptions that may not strictly hold in practice.  We assume that the photon arrival process is Poisson, which does not take into account instrument dead time following the arrival of a photon.  We also assume that the distribution of $z$ does not depend on the arrival time.  This does not take into account possible dependence between energy and the phase of the source (see Fierro et al.~1998) Nonetheless, the form of the statistic $Q_T$ is such that it is sensitive to periodic sources, even when the assumptions upon which it was derived do not strictly hold.

The score test was derived to discriminate between a periodic source and background which is not time varying.  From the nature of the construction, it is clear that a similar test could be derived to take into account a background  intensity  which varies in time in a known way, perhaps for example a known nearby pulsar.

\section{Acknowledgments}
This research was sponsored by the National Science Foundation, Award Number 0507254, and by a VIGRE grant from the National Science Foundation, Award Number  0130526. We thank  Charlotte Wickham and Jeremy Shen for computational assistance and Seth Digel, Patrick Nolan, and Tom Loredo for very helpful conversations.

\section{Appendix}
Here we sketch arguments supporting the assertions about the distribution of the test statistic $Q_T$ under the null and alternative hypotheses. We assume that $\phi(T) \gg 1$ and that $c(t)$ varies slowly and is nonzero over a substantial fraction of $[0,T]$. In particular we assume that  $\Big|\int_0^T \exp(2 \pi i n \phi(t)) c(t) dt \Big|^2$ is negligible compared to $\int_0^T c(t) dt$, which is true, for example, if $c(t)$ is constant.
\subsection{Null distribution}
Let $W(t)=\sum_{j=1}^N w_j\delta(t-t_j)$. Under the null, all events are background and
\begin{eqnarray}
E \frac{1}{\sqrt{T}} A_n  & = & E \frac{1}{\sqrt{T}}\int_0^T e^{2 \pi i n \phi(t)}dW(t)  \\
& = & \frac{(1-\theta) \beta_1 + \theta \zeta_1}{\sqrt{T}} \int_0^T e^{2 \pi i n \phi(t)} \lambda(t)dt \\
& = & \frac{(1-\theta) \beta_1 + \theta \zeta_1}{\sqrt{T}} \mu \int_0^T e^{2 \pi i n \phi(t)} c(t)dt \\
& \simeq & 0
\end{eqnarray}
The approximation holds under the assumptions above about $\phi(t)$ and $c(t)$.
Similarly, the real and imaginary parts of $T^{-1/2}A_n$ are approximately uncorrelated.  To calculate $E |A_n|^2$ we use
\begin{eqnarray}
\label{W20}
E [dW(t)dW(s)] & = & \lambda(t)[(1-\theta)\beta_2 + \theta \zeta_2] \delta(s-t)ds dt  \nonumber \\
& & + \lambda(s)\lambda(t)[\theta^2 \zeta_1^2 + 2 \theta (1-\theta) \zeta_1 \beta_1 + ( 1 - \theta)^2 \beta_1^2] ds dt
\end{eqnarray}
Then
\begin{eqnarray}
E |A_n|^2 & = &  \int_0^T \int_0^T e^{2 \pi i n \phi(t)}e^{-2 \pi i n \phi(s)} E[dW(s) dW(t)] \\
 & = & [(1-\theta) \beta_2 + \theta \zeta_2] \int_0^T \lambda(t) dt + [\theta \zeta_1 + (1- \theta) \beta_1]^2 \left | \int_0^T e^{2 \pi i n \phi(t)} \lambda(t)  dt \right |^2 
\end{eqnarray} 
The first term is dominant.  The limiting chi-squared approximation follows from a central limit theorem argument about the distribution of  the linear statistic $T^{-1/2}A_n$. 

\subsection{Power}

To evaluate $E_KQ_T$ we need to calculate $E[dW(t)dW(s)]$. First, for $s=t$, the event is either with probability $\theta$ from source or with probability $(1-\theta)$ from background.  Thus
\begin{equation}
\label{W211}
E[ dW(s)dW(t)] =  \mu c(t)[\theta \zeta_2 \gamma(\phi(t)) + (1-\theta)\beta_2]dt, ~~ s=t
\end{equation}
For $s \neq t$ there are three possibilities: both events are from source, both are from background, or one is from source and one is from background.
\begin{eqnarray}
\label{W212}
E[dW(t)dW(s)] &  = & \mu^2 c(s)c(t)  [
  \theta^2 \zeta_1^2 \gamma(\phi(t))\gamma(\phi(s))  +
(1 - \theta)^2 \beta_1^2 + \nonumber \\
& & \theta(1-\theta)\zeta_1 \beta_1 (\gamma(\phi(s)) + \gamma(\phi(t))  
]dsdt, ~~s \neq t
\end{eqnarray}

We initially assume that the phase $\phi(t)$ is properly specified, i.e. that $f$ and $\dot{f}$ are identified.     $E |A_n|^2$ contains contributions of all the terms in (\ref{W211}) and (\ref{W212}).  Some analysis shows that the  leading order  comes from the first  term in (\ref{W212}), leading to
\begin{eqnarray}
\mu^2 \theta^2 \left| \int e^{-2 \pi i n \phi(t)} c(t) \gamma(\phi(t))  dt  \right|^2& = & 
\mu^2 \theta^2 \left| \sum_k \int \gamma_k c(t) e^{-2 \pi i n \phi(t)} e^{2 \pi i k \phi(t)} dt \right|^2 \\
 & \simeq & \mu^2 \theta^2 |\gamma_n|^2 \Big[ \int_0^T c(t) dt \Big]^2
\end{eqnarray}
Thus, under the alternative
\begin{equation}
\label{ }
E_K Q_T \simeq \mu^2 \theta^2 \zeta_1^2 \frac{[\int_0^T c(t) dt]^2}{T} \sum_{n \neq 0}
|\gamma_n|^2 |\alpha_n|^2
\end{equation}
The approximation for frequency misspecification, $\Delta \neq 0$,  follows from Taylor series expansions, noting that the first derivatives vanish at $\Delta=0$, since that point is a maximum.

\end{document}